\documentclass[iop]{emulateapj}

\slugcomment{}
\shorttitle{Non-thermal X-rays from radio lobes of Cygnus A}
\shortauthors{Yaji et al.}

\begin{document}
\title{Evidence of non-thermal X-ray emission from radio lobes of Cygnus A}
\author{Y. Yaji\altaffilmark{1},
  M. S. Tashiro\altaffilmark{1},
  N. Isobe\altaffilmark{2},  
  M. Kino\altaffilmark{3},
  K. Asada\altaffilmark{4},
  H. Nagai\altaffilmark{3},
  S. Koyama\altaffilmark{3,5},
  and M. Kusunose\altaffilmark{6}}

\email{yaji@heal.phy.saitama-u.ac.jp}
\altaffiltext{1}{Department of Physics, Saitama University,
                 Shimo-Okubo, Sukura-ku, Saitama 338-8570, Japan}
\altaffiltext{2}{Department of Astronomy, Kyoto University,
                  Sakyo-ku, Kyoto 606-8502, Japan}
\altaffiltext{3}{National Astronomical Observatory of Japan
                 2-21-1 Osawa, Mitaka, Tokyo 181-8588, Japan}
\altaffiltext{4}{Academia Sinica Institute of Astronomy and Astrophysics,
                 Taipei, Taiwan}
\altaffiltext{5}{Department of Astronomy, Graduate School of Science, 
                  The University of Tokyo, 7-3-1 Hongo, Bunkyo-ku, Tokyo 113-0033 }
\altaffiltext{6}{Department of Physics, School of Science and Technology, 
                 Kwansei Gakuin University, Sanda, Hyogo 669-1337, Japan}

%%%%%%%%%%%%
% Abstract %
%%%%%%%%%%%%
\begin{abstract}
Using deep $\it Chandra$ ACIS observation data for Cygnus A,
we report evidence of non-thermal X-ray emission from radio
lobes surrounded by a rich intra-cluster medium (ICM).
The diffuse X-ray emission, which are associated with the eastern 
and western radio lobes, were observed in a 0.7--7 keV $\it Chandra$ 
ACIS image. The lobe spectra are reproduced with not only a 
single-temperature Mekal model, such as that of the surrounding ICM 
component, but also an additional power-law (PL) model.
The X-ray flux densities of PL components for the eastern and 
western lobes at 1 keV are derived as $77.7^{+28.9}_{-31.9}$ nJy and 
$52.4^{+42.9}_{-42.4}$ nJy, respectively, and the photon indices are 
$1.69^{+0.07}_{-0.13}$ and $1.84^{+2.90}_{-0.12}$, respectively.
The non-thermal component is considered to be produced via the inverse 
Compton (IC) process, as is often seen in the X-ray emission from radio 
lobes. From a re-analysis of radio observation data, the multiwavelength
spectra strongly suggest that the seed photon source of the IC X-rays
includes both cosmic microwave background radiation and synchrotron radiation
from the lobes. The derived parameters indicate significant dominance of 
the electron energy density over the magnetic field energy density
in the Cygnus A lobes under the rich ICM environment. 
\end{abstract}


%%%%%%%%%%%%%
% Section 1 %
%%%%%%%%%%%%%
\section{Introduction}
Radio lobes, in which jets release a fraction of the kinetic energy 
originating from active galactic nuclei (AGNs), store enormous amounts
of energy as relativistic electrons and magnetic fields. Relativistic 
electrons in the lobes emit synchrotron radiation (SR) at radio frequencies 
and boost the seed photons into the X-ray and $\gamma$-ray ranges via the 
inverse Compton (IC) process. Candidates for seed photons are cosmic 
microwave background (CMB) photons (e.g., Harris \& Grindlay 1979), infrared 
(IR) photons from the host AGN (Brunetti et al. 1997) and SR photons emitted 
in the lobes. If the seed photon sources can be identified, 
the energy densities of relativistic electrons ($u_{\rm e}$) 
and magnetic fields ($u_{\rm m}$) can be determined from a comparison 
of the SR and IC fluxes, respectively. These energy densities can provide 
important clues regarding the energy of astrophysical jets and the evolution 
of radio galaxies.

%%==========%%
%% Figure 1 %%
%%==========%%
\begin{figure*}[t]
\centerline{
\includegraphics[angle=0,width=7.55cm]{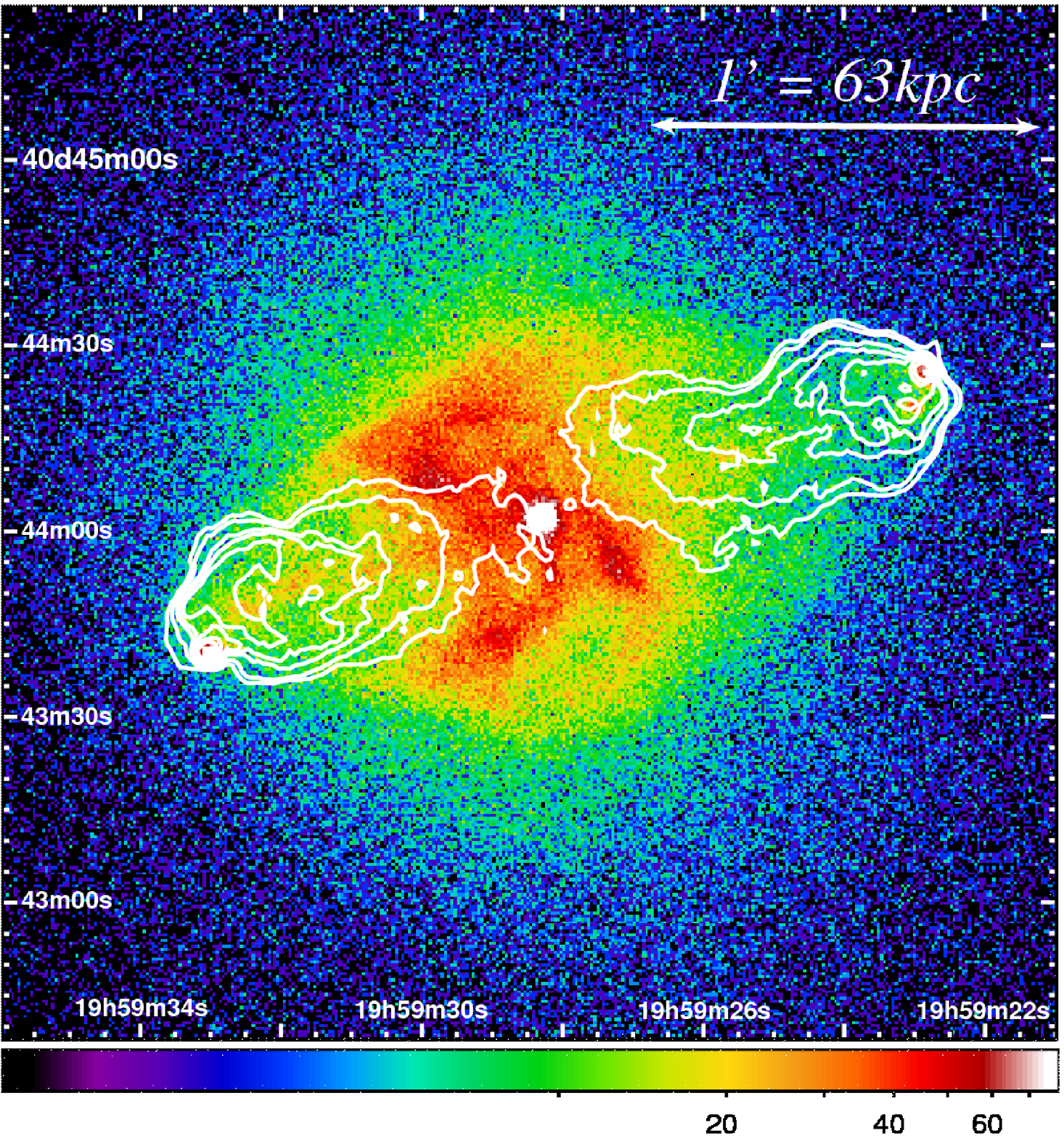}
\includegraphics[angle=0,width=7.55cm]{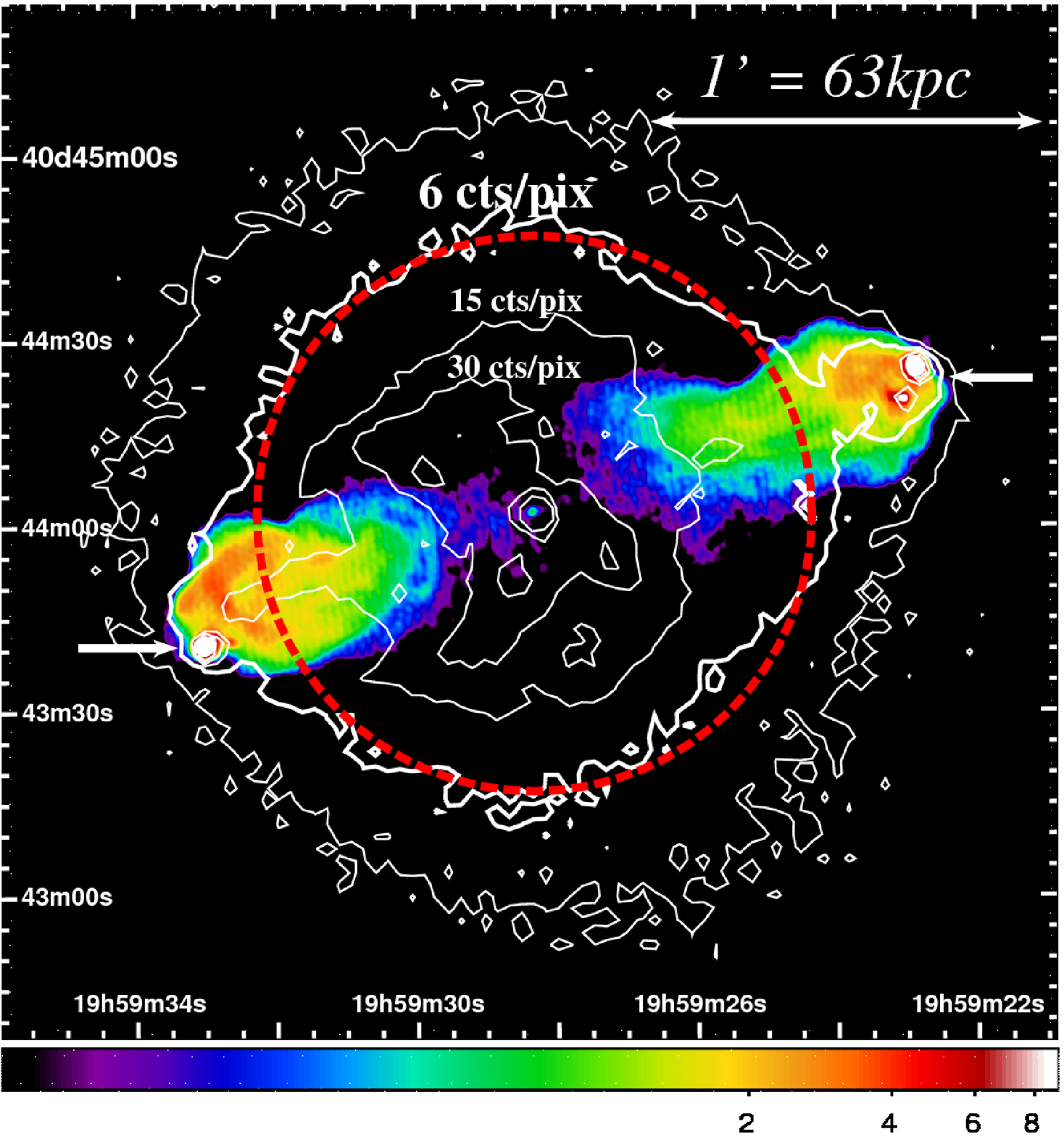}}
\caption{
$\it Left$: The co-added raw 0.7--7 keV ACIS image of Cygnus A.
Color scale of the image shows the photon counts for each pixel;
scale bar is shown below. White contours represent the radio strength 
of the 1.3 GHz band observed by the VLA, the contours levels are 
0.053, 0.2, 0.8, 1.2, 2, 5 and 10 $\rm Jy~beam^{-1}$ for a beam size 
of 1.19$\arcmin$$\times$1.12$\arcsec$. $\it Right$:
ACIS contours image of Cygnus A at 0.7--7 keV, superposed on 
the 1.3 GHz VLA image of the $\rm Jy~beam^{-1}$ for each pixel;
scale bar is shown below. Five contours represent X-ray brightness
of 3, 6, 15, 30, 50 and 100 counts per pixel.
}
\label{X-rayImage}
\end{figure*}

Due to the faintness of the IC X-rays emitted from the lobes, the 
objects located at the edge of the cluster of galaxies have been 
targeted, owing to the poor radiation from the intra-cluster medium (ICM).
In the past 20 years, around 30 objects have been observed with $ASCA$ 
and $ROSAT$ (e.g., Kaneda et al. 1995; Feigelson et al. 1995; Tashiro 
et al. 1998, 2001), as well as with the $\it Chandra~X$-$\it ray~ 
observatory$, $\it XMM$-$\it Newton$ and $\it Suzaku$ (e.g., Burunetti 
et al. 2001; Isobe et al. 2002, 2005; Hardcastle et al. 2002; Grandi et al. 
2003; Comastri et al. 2003; Croston et al. 2004; Kataoka \& Stawarz 2005; 
Tashiro et al. 2009). In most cases, the seed photons were determined to be
CMB photons. The measured IC X-ray flux from the lobes often requires that 
$u_{\rm e}$ is considerably greater than $u_{\rm m}$ (e.g., Tashiro et al. 
1998; Isobe et al. 2002), as well as that the magnetic field ($B_{\rm IC}$) 
is smaller than the magnetic field estimated under equipartition ($B_{\rm eq}$), 
$B_{\rm IC}/B_{\rm eq}=0.1$--1 (e.g., Croston et al. 2005). Against this 
background, it is of great interest to investigate the energy balance between 
the relativistic electrons and the magnetic fields under ICM pressure 
in order to argue the energetics from the nuclei to 
inter-galactic space.

In this paper, we present an examination of the diffuse lobe emission
from one of the brightest radio lobe objects surrounded by ICM, Cygnus A.
Cygnus A is a well-known FR II (Fanaroff \& Riley 1974) radio galaxy 
with an elliptical host. The radio images show symmetrical double-lobe
morphology with an extremely high radio flux density $S_{\rm SR}=1598$ 
Jy at 1.3 GHz (B{\^i}rzan et al. 2004), which makes it the brightest radio
 galaxy in the observable sky. Measuring radio fluxes between 151 MHz and 
5000 MHz, Carilli et al. (1991) found spectral breaks whose break frequencies
vary with position on the lobes. The spectral energy index below and above 
the break is 0.7 and 2, respectively. X-ray observations show diffuse X-ray
emission, which is considered to originate from ICM (e.g., Smith et al. 2002),
as well as a cavity corresponding to the radio lobe (e.g., Wilson et al. 2006).
In addition to the thermal emissions, we examined non-thermal X-ray
emissions from the lobes of Cygnus A utilizing the excellent spatial
resolution of {\it Chandra}.

The structure of this paper is as follows. We describe the archived 
$\it Chandra$ observation data on Cygnus A and the results of careful 
X-ray analysis in $\S~2$ and $\S~3$. Following the presentation of the 
results, namely, the suggestion of non-thermal X-ray emissions from the 
lobes, we report the spectral energy distribution of the Cygnus A lobes, 
as determined from radio and X-ray data, and estimate the emission of seed 
photons and the physical parameters of the lobes in $\S~4$. Finally, we 
summarize these results in the last section. Throughout this paper,
we adopt a cosmology with $H_0=71$ km s$^{-1}$ Mpc$^{-1}$,
$\Omega_{\rm M}=0.27$ and $\Omega_{\Lambda}=0.73$ (Komatsu et al. 2009),
where $1\arcmin$ corresponds to 63 kpc at the red shift
$\it z$=0.0562 (Stockton et  al. 1994) of Cygnus A.

%%%%%%%%%%%%%
% Section 2 %
%%%%%%%%%%%%%
\section{X-ray observation and data reduction}
Cygnus A has been observed with the Advanced CCD Imaging 
Spectrometer (ACIS) detector on the $\it Chandra~X$-$Ray~Observatory$ 
on nine occasions in the full frame mode. Seven out of nine the 
observations were performed with ACIS-I1 front-illuminated (FI) CCDs,
one with ACIS-I3 FI CCDs and the rest with ACIS-S3 back-illuminated (BI)
CCDs. The total exposure of ACIS-I1, ACIS-S3 and ACIS-I3 is 172.2 ks,
34.7 ks and 29.7 ks, respectively. Table \ref{obs_log} summarizes the 
nine ACIS observations of Cygnus A. These observations were performed with 
the default frame time of 3.2 s using the FAINT (ACIS-S3) and VFAINT 
(ACIS-I1 and I3) format.

%%--------%%
%% Table1 %%
%%--------%%
\begin{deluxetable}{lccccc}
\tablecaption{Observation log of $Chandra$/ACIS\label{obs_log}}
\tablewidth{0pt}
\tablehead{
\colhead{OBS-ID}	 &
\colhead{~~~~~~~~}	 &
\colhead{Instrument}	 &
\colhead{Date } &	 
\colhead{Exposure (ks)}	 
}
\startdata
0360 && ACIS-S3 & 2000.05.21 &  34.7\\
6225 && ACIS-I1 &2005.02.15 &  25.8\\
5831 & & ACIS-I1 &2005.02.16 &  51.1\\
6226 & & ACIS-I1 &2005.02.19 &  25.4\\
6250 & & ACIS-I1 &2005.02.21 &  7.0\\
5830 & & ACIS-I1 &2005.02.22 &  23.5\\ 
6229 & & ACIS-I1 &2005.02.23  & 23.4\\  
6228 & & ACIS-I1 &2005.02.25  &  16.0\\
6252 && ACIS-I3 & 2005.09.07 & 29.7\\
total & & -- & --&172.2 
\enddata
\end{deluxetable}

The data were reduced using the CIAO version 4.1 
software package\footnote{http://cxc.harvard.edu/ciao/},
and we performed the data analysis using HEASOFT version 6.8.
We reprocessed all the data by following standard procedures 
in order to create new ``level-2'' event files by utilizing CALDB 
version 4.1.2. We applied the latest gain and charge transfer inefficiency
corrections and a new bad pixel file created with \texttt{acis\_run\_hotpix}.
We generated a clean data set by selecting standard grades (0, 2, 3, 4 and 6). 
After removing the point X-ray sources within the field of view of ACIS and 
the Cygnus A region, we produced a light curve covering the entire CCD chip.
Using \texttt{lc\_sigma\_chip}, we excluded high background time regions whose threshold was more than 3 $\sigma$ above the mean. 
 
Thus, we obtained good exposure of 167.6 ks for ACIS-I1, 29.7 ks for 
ACIS-I3 and 34.5 ks for ACIS-S3.

%%==========%%
%% Figure 2 %%
%%==========%%
\begin{figure*}[t]
\centerline{
\includegraphics[angle=0,width=7.55cm]{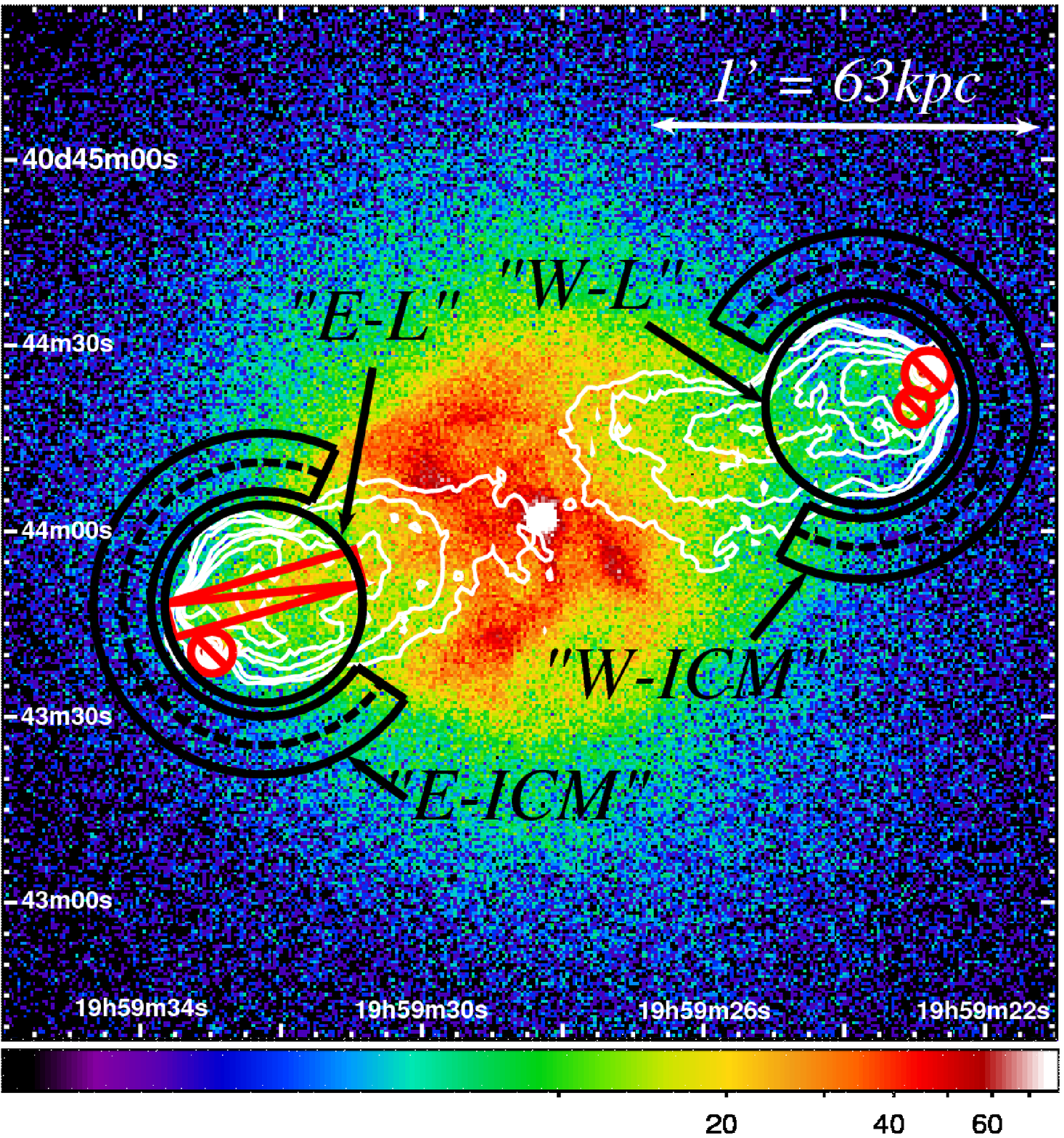}
\includegraphics[angle=0,width=7.55cm]{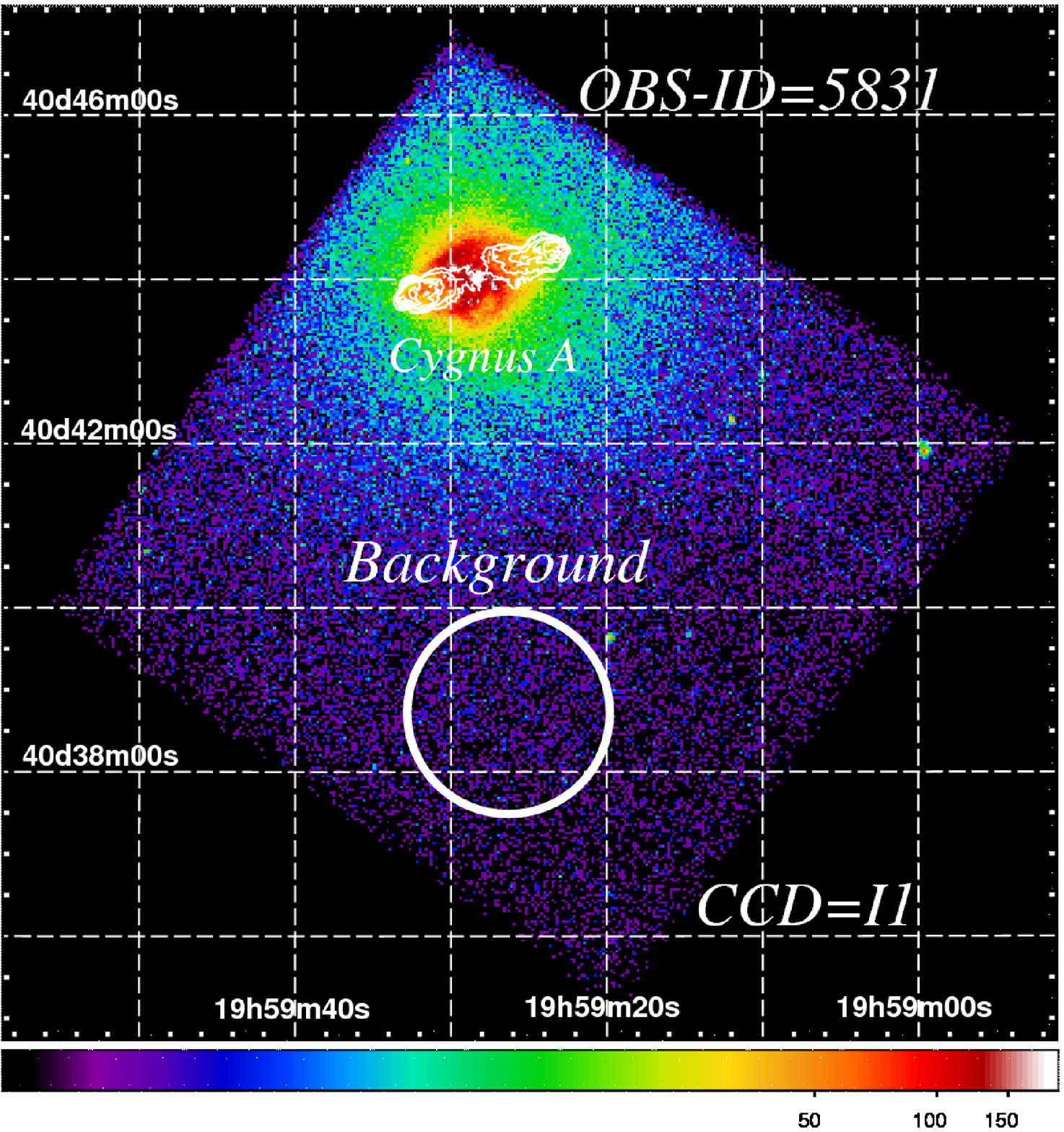}}
\caption{
$\it Left$:
ACIS X-ray image in the 0.7--7 keV range.
The 1.3 GHz VLA contour is overplotted. 
The integration region for spectral analysis is shown.
The areas enclosed by black solid-line circles are the ``lobe regions''.
The red solid-line circle and rectangle represent excluded hotspots and 
jet regions. The areas enclosed by solid-line concentric annuli are the 
regions surrounding the lobes (see text). $\it Right$:
Raw ACIS--I1 data of obsid=5831 X--ray image, binned into $4 \times 4$ 
ACIS pixels. The 1.3 GHz VLA contour is overplotted.
The area enclosed by the solid-line circle denotes the background region.
}
\label{X-rayImage_IntgrationRegion}
\end{figure*}

%%%%%%%%%%%%%
% Section 3 %
%%%%%%%%%%%%%
\section{Results of X-ray analysis}

\subsection{X-ray image and selection of integration region}
The left panel of Fig. \ref{X-rayImage} shows the 0.7--7 keV 
$\it Chandra$ image composed of nine co-added data sets with 
\texttt{merge\_all} from CIAO in gray scale, on which 1.3 GHz 
radio contours are overplotted. Details of the radio data are 
given in \S~4. The AGN nucleus is observed at 
$\alpha(2000)=19^{h}59^{m}28.3^{s}$, 
$\delta(2000)=+40\arcdeg 44\arcmin 02\arcsec$.
The hotspots are located in the eastern and western lobe.
The distance between the two hotspots is $\sim120\arcsec$, 
which corresponds to the projected distance of about 126 kpc. 
The western lobe is approaching and the eastern lobe is receding 
from us (Perley et al. 1984). X-ray emission that extends to the 
vicinity of the nucleus is explained by Smith et al. (2002) as the 
ICM of $\it kT \sim$ 4--9 keV. The right panel of Fig. \ref{X-rayImage}
shows the 1.3 GHz VLA image in gray scale on which X-ray brightness 
contours (solid line) are overplotted. From the X-ray contours of 15 
counts per pixel, faint X-ray emission can be seen from the jet located 
to the east of the nucleus ($\alpha(2000)=19^{h}59^{m}31.0^{s}$, 
$\delta(2000)=+40\arcdeg 43\arcmin 52\arcsec$),
and the eastern hotspot (Steenbrugge et al. 2008).
It appears that the X-ray contours at 15 counts per pixel 
(thin line) form a ``cavity'' in the ICM avoiding the 1.3 GHz lobe.
On the other hand, beyond the dashed circle at $45\arcsec$ 
from the nucleus, the contours at 6 counts per pixel 
(thick line) show an extended structure in the direction of 
the 1.3 GHz lobe on the east and west sides 
(arrows in the right panel of Fig. \ref{X-rayImage}).
These extended structures are apparently associated with the lobes,
although the X-rays from the lobes are thought to be contaminated with 
X-rays from the foreground and background ICM. 

In order to evaluate the extended emission along the lobes,
we accumulated the X-ray spectra from the regions
shown in Fig. \ref{X-rayImage_IntgrationRegion} ($\it left$). 
The regions ``Eastern--Lobe'' (``E--L'') and ``Western--Lobe'' 
(``W--L'') are marked with circles with a radius of $16 \arcsec$ 
(16.8 kpc), corresponding to the 1.3 GHz contours.
However, in order to avoid X-ray emissions from hotspots and jets
(Steenbrugge et al. 2008), we excluded hotspots and jet regions. 
Moreover, to investigate the ICM in the foreground and background 
of the lobes, we accumulated X-ray spectra from the ICM surrounding 
the lobes. We set concentric annuli around the lobe, with inner and 
outer radii of 18.2$\arcsec$ and 27.6$\arcsec$, respectively, which 
are denoted as ``Eastern--ICM'' (``E--ICM'')  and  ``Western--ICM'' 
(``W--ICM''), as shown in Fig. \ref{X-rayImage_IntgrationRegion} 
($\it left$). The region facing the nucleus is excluded so as to avoid 
contamination from the lobes and jets.

The background spectrum is estimated in another circular region,
avoiding bright ICM emission as well as point sources.
The background spectra are accumulated from the same circular
 region whose radius is 74$\arcsec$ (77.7 kpc) and the center is placed at 
 ($\alpha(2000)=19^{h}59^{m}29.2^{s}$,
 $\delta(2000)=+40\arcdeg 38\arcmin 29\arcsec$)
 for the ACIS-I1 and ACIS-S3.
 Because of the different roll angle at the ACIS-I3 observation,
 we cannot choose the same region for the background, but had to set the
 center at ($\alpha(2000)=19^{h}59^{m}52.3^{s}$,
 $\delta(2000)=+40\arcdeg 46\arcmin 38\arcsec$) 
with a radius same to the background of the ACIS-I1 and ACIS-S3.
 Although the two regions may contain different unresolved sources, we
 confirmed there is no difference in the obtained background spectra
In Fig. \ref{X-rayImage_IntgrationRegion} ($\it right$),
the background region of obsid=5831 is shown as an example.

%%==========%%
%% Figure 3 %%
%%==========%%
\begin{figure*}
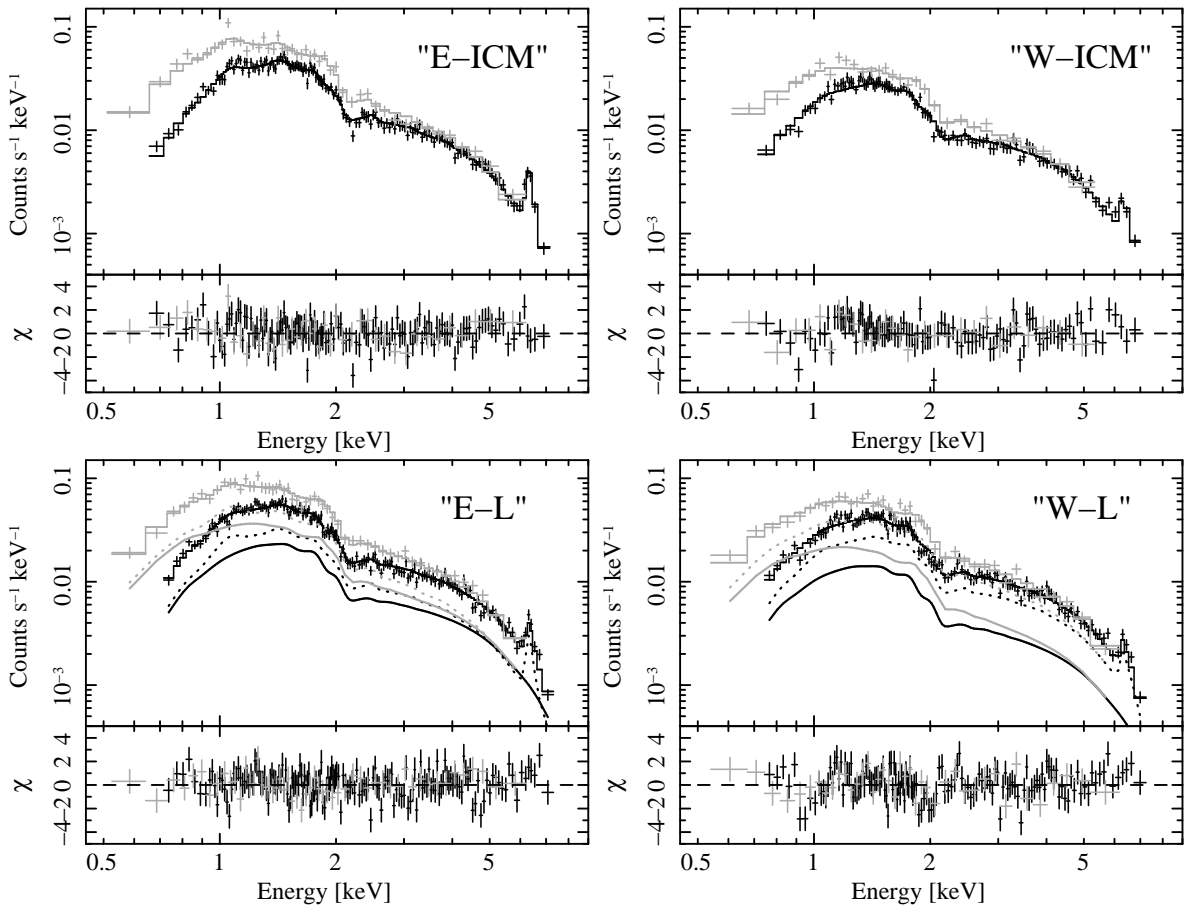
%[!b]
\centerline{
\includegraphics[angle=270,width=7.8cm]{fig3-1.eps}
\includegraphics[angle=270,width=7.8cm]{fig3-2.eps}}
\centerline{
\includegraphics[angle=270,width=7.8cm]{fig3-3.eps}
\includegraphics[angle=270,width=7.8cm]{fig3-4.eps}}
\caption{
$\it top$:
Background-subtracted 0.7--7 keV ACIS-FI (black) and 
0.5--6 keV ACIS-BI (gray) spectra of the regions surrounding
the eastern (left) and western (right) lobes. Histogram is a 
best-fit Mekal model spectrum absorbed by galactic $N_{\rm H}$.
The lower panel shows the residuals of the absorbed Mekal model fit.  
$\it bottom$: Background-subtracted 0.7--7 keV ACIS-FI (black) and 
0.5--6 keV ACIS-BI (gray) spectra of the eastern (left) and western 
(right) lobes. The histogram with the solid line is a best-fit (PL+Mekal) model.
The included PL and Mekal components are indicated by solid and dotted lines. 
Both model spectra are modified by galactic absorption.
Lower panel shows the residuals of the absorbed (Mekal+PL) model.  
}
\label{X-raySpectrumMekal}
\end{figure*}

\subsection{X-ray spectrum of ICM surrounding the lobe}
We first investigated the possible temperature gradient of
the ICM in the ``E--ICM'' and ``W--ICM'' regions. 
For that purpose, we divided each region in two, as shown by 
the dashed lines in Figure \ref{X-rayImage_IntgrationRegion} 
($\it left$): the inner regions are ``E--ICM1'' and ``W--ICM1'', and 
the outer regions are ``E--ICM2'' and ``W--ICM2''. Spectra were extracted 
from all of nine data sets using \texttt{specextract} from CIAO. We added
the spectra acquired with FI CCDs (ACIS-I1 and ACIS-I3) using \texttt{mathpha}
from FTOOL and performed joint fitting with the spectrum acquired with BI 
CCDs (ACIS-S3). In the following, these spectra are referred to as 
ACIS-FI (ACIS-I1 and ACIS-I3) and ACIS-BI (ACIS-S3). Furthermore, we introduced 
a constant factor for the relative normalization of ACIS-FI and ACIS-BI.
Then, we introduce a single-temperature Mekal model (Mewe et al. 1995) to 
estimate the ICM component absorbed by Galactic $N_{\rm H}$ (fixed at $3.5 
\times 10^{21}\rm ~cm^{-2}$; Dickey \& Lockman 1990). The solar abundances 
of Anders \& Grevesse (1989) were used throughout. The response matrix files
(rmf) and the auxiliary response file (arf) for the model
were produced from observation data with \texttt{specextract} from CIAO, 
and then weighted by their exposure and added with \texttt{addrmf} 
and \texttt{addarf} from FTOOL. Each energy bin was set to include at least 
80 counts ($\sim9\sigma$/bin). In consideration of the decrease in source 
photons, the quantum efficiency and the increase in the background that is 
intrinsic to the detector in high and low energy bands, we selected the more 
reliable 0.7--7 keV range for ACIS-I and 0.5--6 keV for ACIS-S to maximize 
the signal-to-noise ratio. The best-fit spectral parameters are listed in 
Table \ref{lobe_ambient}, and we obtained an acceptable fit.
The values for the inner and outer regions are in fairly good agreement. 
Thus, significant $kT$ and $Z/Z_{\odot}$ gradients were not observed in 
the ICM surrounding the lobe within a 90\% confidence level. Next, we 
analyzed ``E--ICM'' and ``W--ICM''. Figure \ref{X-raySpectrumMekal} (top) 
presents the background-subtracted ACIS-I 0.7--7 keV and ACIS-S 0.5--6 keV 
spectra of the ``E--ICM'' and ``W--ICM'' regions. The best-fit spectral 
parameters are listed in Table \ref{lobe_ambient}. We obtained an acceptable 
fit with $\chi^2/dof$=190.13/185 (``E-ICM") and 135.76/130 (``W-ICM") 
($P(\chi)=0.382$ and $0.347$, respectively). Therefore, we adopted the 
values of $kT$ and $Z/Z_{\odot}$ derived in this analysis as the spectral 
parameters for the ICM foreground and background of the lobes.

%%--------%%
%% Table2 %%
%%--------%%
\begin{deluxetable*}{lcccccc}
\tablecaption{
Best-fit spectral parameters of ICM surrounding the Cygnus A lobes
\label{lobe_ambient}}
\tablewidth{0pt}
\tablehead{
\colhead{Region}	 &
\colhead{$N_{\rm H}$\tablenotemark{a}}	 &
\colhead{$kT$ (keV)} &
\colhead{ $Z/Z_{\odot}$\tablenotemark{b}}	 &
\colhead{ Norm\tablenotemark{c}} &	
\colhead{Constant\tablenotemark{e}}	 &
\colhead{$\rm \chi^2$/dof}	 
}
\startdata
E--ICM1  & 0.35\tablenotemark{d}  &$5.38^{+0.43}_{-0.42}$  &  $0.62^{+0.20}_{-0.19}$  
&  $4.06^{+0.25}_{-0.17}$    &1.05$^{+0.07}_{-0.06}$  &101.63/91\\% (0.2094)\\
E--ICM2  & 0.35\tablenotemark{d}  &$4.81^{+0.40}_{-0.31}$  &  $0.65^{+0.19}_{-0.18}$  
&  $4.39^{+0.24}_{-0.23}$    &  1.10$\pm 0.06$&95.47/98\\% (0.5534)\\
E--ICM  & 0.35\tablenotemark{d}  &$5.11 \pm 0.28$  &  $0.63 \pm 0.13$  
&  $8.50^{+0.32}_{-0.31}$    & 1.08$\pm 0.04$ &190.13/185\\% (0.3825)\\
\hline%-----------------------------------------------------------------------
W--ICM1  & 0.35\tablenotemark{d}  &$8.07^{+1.33}_{-1.07}$  &  $0.49^{+0.33}_{-0.30}$  
&  $2.51 \pm 0.17$   & 1.08$\pm 0.08$ &62.75/62\\% (0.4496)\\
W--ICM2  & 0.35\tablenotemark{d}  &$6.87^{+0.85}_{-0.79}$  &  $0.57^{+0.27}_{-0.24}$  
&  $3.00 \pm 0.18$   & 0.97$\pm 0.07$ &73.96/73\\% (0.4465)\\
W--ICM  & 0.35\tablenotemark{d}  &$7.25^{+0.77}_{-0.60}$  &  $0.48^{+0.18}_{-0.17}$  
&  $5.56\pm0.23$   & 1.01$^{+0.06}_{-0.05}$ &135.76/130
\enddata
\tablenotetext{a}{Photoelectric absorption column density in units of 10$^{22}$ H atom cm$^{-2}$.}
\tablenotetext{b}{Metal abundance normalized by the solar abundance.}
\tablenotetext{c}{Emission measure in units of $10^{-10} \int n_{\rm e} n_{\rm H}dV / 4 \pi \{\rm d_{\rm A} (1+z)\}^2$.}
\tablenotetext{d}{Fixed at the Galactic line-of-sight value.}
\tablenotetext{e}{Normalization factor of the ACIS-BI relative to that of the ACIS-FI}
\end{deluxetable*}

%%--------%%
%% Table3 %%
%%--------%%
\begin{deluxetable*}{lccccccccc}
\tablecaption{Results for four models fit to the ACIS spectrum of the lobe region \label{Eresult}}
\tablewidth{0pt}
\tablehead{
\colhead{}	 &
\colhead{}	 &
\colhead{$kT_1$} &
\colhead{}	 &
\colhead{$kT_2$(keV)} &	
\colhead{}	 &
\colhead{}	 &
\colhead{Norm$_2$\tablenotemark{b} } &
\colhead{}	 &
\colhead{}	 \\
Region
& M\tablenotemark{a}
&  (keV)  
&  $Z_1/Z_{\odot}$
& Norm$_1$\tablenotemark{b}
&  $\Gamma_{\rm X}$
& $Z_2/Z_{\odot}$
& $S_{1\rm keV}$\tablenotemark{c} (nJy)
& Constant\tablenotemark{e}
& $\chi^2$/dof
}
\startdata
 E-L& (a)  &  5.11\tablenotemark{d} (fix)   &  0.63\tablenotemark{d} (fix)     &   $10.36^{+0.17}_{-0.19}$ &  --                                          & --                                                & --                                          &1.08$\pm0.04$  &286.03/217 \\
       & (b)  &  $5.84^{+0.36}_{-0.30}$         &  $0.51^{+0.12}_{-0.11}$           &   10.48$^{+0.33}_{-0.32}$ &  --                                          & --                                               & --                                           &1.09$\pm0.04$& 244.83/215 \\
       & (c)  &  5.11\tablenotemark{d} (fix)    &  0.63\tablenotemark{d} (fix)    &   8.33$^{+0.30}_{-5.63}$    & 15.32$^{+\infty}_{-9.20}$  &  0.63\tablenotemark{d} (fix) & 2.01$^{+5.49}_{-0.91}$  &1.09$\pm0.04$& 244.03/215\\
       & (d)  &  5.11\tablenotemark{d} (fix)    &  0.63\tablenotemark{d} (fix)    &   5.84$^{+1.51}_{-1.54}$   & $1.69^{+0.07}_{-0.13}$     &  --                                              & $77.7^{+28.9}_{-31.9}$  &1.08$\pm0.04$& 224.06/215\\
\hline%-----------------------------------------------------------------------
W-L& (a) & 7.25\tablenotemark{d} (fix)  &  0.48\tablenotemark{d} (fix)    & 7.92$^{+0.16}_{-0.15}$ &   -- & -- & -- &1.03$^{+0.05}_{-0.04}$ & 273.18/178 \\
       &  (b) & 6.55$^{+0.49}_{-0.47}$  &  0.43$\pm0.13$  & 8.08$^{+0.28}_{-0.27}$ &   -- & -- & -- 
      &1.03$^{+0.04}_{-0.05}$& 262.83/176 \\
      &(c) & 7.25\tablenotemark{d} (fix)  &  0.48\tablenotemark{d} (fix)    & 7.87$^{+0.15}_{-0.16}$ & 0.22$^{+0.46}_{-0.08}$ & 0.48\tablenotemark{d} (fix) & 0.85$^{+3.45}_{-0.75}$ &1.02$^{+0.04}_{-0.05}$& 255.54/176 \\
      &(d) & 7.25\tablenotemark{d} (fix)  &  0.48\tablenotemark{d} (fix)    & 5.31$^{+2.49}_{-2.01}$ & $1.84^{+2.90}_{-0.12}$ & --   & $52.4^{+42.9}_{-42.4}$&1.02$^{+0.04}_{-0.05}$ & 241.29/176
\enddata
\tablenotetext{a}{Models (a) wabs$\times$Mekal (fix), (b) wabs$\times$Mekal (free), (c) wabs$\times$(Mekal+Mekal), (d) wabs$\times$(Mekal+PL) (see text).} 
\tablenotetext{b}{Emission measure in unit of $10^{-10} \int n_{\rm e} n_{\rm H}dV / 4 \pi \{\rm d_{\rm A} (1+z)\}^2$.}
\tablenotetext{c}{Flux density at 1 keV.}
\tablenotetext{d}{The best-fit parameters for $kT$ and $Z/Z_{\odot}$ obtained from ``E--ICM'' or ``W--ICM'' are employed as fixed parameters.}
\tablenotetext{e}{Normalization factor of the ACIS-BI relative to that of the ACIS-FI}
\end{deluxetable*}

\subsection{X-ray spectrum of lobe region}
Figure \ref{X-raySpectrumMekal} (bottom left) shows the 
background-subtracted ACIS-I 0.7--7 keV and ACIS-S 0.5--6 keV 
spectra from the ``E--L'' region. {\bf Model (a)}: Firstly, we 
evaluated the spectrum with a single-temperature Mekal model 
whose $kT$ and $Z/Z_{\odot}$ were fixed as the values for 
the ICM surrounding the lobe regions ``E--ICM'' (\S 3.2). 
The obtained value of $\chi^2/dof=286.0/217$ ($P(\chi)=0.001$)
indicates that the single-temperature model is unacceptable.
{\bf Model (b)}: Secondly, we refit the spectrum with the Mekal 
model whose $kT$ and $Z/Z_{\odot}$ were freed.
The best-fit parameters are listed in Table \ref{Eresult}.
Again, we obtained a unacceptable fit with $\chi^2/dof=244.8/215$ 
($P(\chi)=0.079$), and the evaluated $kT$ was significant larger 
than that of the ICM surrounding the lobe. Unlike the ``E--ICM'' 
spectrum, the ``E--L'' spectrum cannot be described with a single-temperature
 thermal emission model. In order to determine the other X-ray 
emission mechanism, we estimated the spectrum with additional models.
{\bf Model (c)}: We added second Mekal model. The values of $kT$ and 
$Z/Z_{\odot}$ for one of the two components were fixed as the average 
values for the ICM surrounding the lobe, in order to represent the ICM in 
the foreground and background of the lobe, but the values for the second 
component were freed to fit the spectrum. The best-fit parameters are 
presented in Table \ref{Eresult}, but the resultant value of 
$\chi^2/dof=244.03/215$ ($P(\chi)=0.085$) is not significantly improved.
{\bf Model (d)}: Finally, we added a power-law (PL) model instead of the 
additional Mekal model in (c). The best-fit parameters are shown in 
Table \ref{Eresult}. The improved value of $\chi^2/ \rm dof = 224.06/215$ 
($P(\chi)=0.342$) indicates that this fit is acceptable.
Figure \ref{X-raySpectrumMekal} (bottom left), shows the histogram 
of the PL+Mekal model (upper panel) and the residual (lower panel).
The best-fit photon index and flux density at 
1 keV for the PL are $\Gamma$=$(1.69\pm0.06)^{+0.03}_{-0.12}$ 
and $S_{\rm 1keV}$=$(77.7^{+28.3}_{-28.7})^{+6.0}_{-14.0}~ \rm nJy$, 
respectively. The first error for each parameter indicates the 90\% 
confidence level including the error arising from normalization of the 
Mekal model and the systematic error between ACIS-FI and ACIS-BI.
The second error indicates the systematic error at the 90\% confidence 
level for both $kT$ and $Z/Z_{\odot}$ from the evaluation of the ``E--ICM'' 
spectrum (\S 3.2). The resultant errors are shown in Table \ref{Eresult}, 
and the PL flux is not a 3-$\sigma$ detection. As an interpretation of 
non-thermal emission, the PL model naturally implies a lobe source.

The derived high temperature for the second Mekal component in Model (c) 
possibly suggests shock-heating in the ICM or additional high-temperature 
plasma in the lobe. Although a shock-heated shell has been detected in 
Cygnus A (Wilson et al. 2006),  the shock-heated shell region is not included 
in the spectrum-integrated regions in the present study. Furthermore, the 
derived thermal electron density of $(1.37^{+2.42}_{-0.99})\times 10^{-2}
~\rm cm^{-3}$ of the eastern lobe, where we assumed that thermal plasma 
volume is a sphere of radius 16.8 kpc in the integrated region, 
is larger by nearly two orders of magnitude than the thermal electron density 
of $<10^{-4}~\rm cm^{-3}$ reported by Dreher et al. (1987) and Carilli et al.
(1998), based on their radio rotation measurements from the region. However, 
this limit can be two orders of magnitude larger assuming a tangled magnetic 
field. Therefore, the thermal interpretation is still a possibility.

Figure \ref{X-raySpectrumMekal} (the bottom right) shows the 
background-subtracted ACIS-I 0.7--7 keV and ACIS-S 0.5--6 keV spectra from 
the ``W--L'' region. Moreover, in Table \ref{Eresult}, each best-fit parameter 
is listed for the spectra in the ``W--L'' region, which is evaluated in the 
same manner as the ``E--L'' region. Although the X-ray brightness of the 
``W--L'' region is half that of the ``E--L'' region and the temperature of 
the ICM surrounding the western lobe is fairly high, the improvement is not 
clear. Among the four models, Model (d) provides the best result, $\chi^2/ 
\rm dof = 241.29/176$ ($P(\chi)=0.01$), as in the case of the ``E--L'' 
spectrum. The best-fit photon index and flux density at 1 keV for the PL are 
$\Gamma$=$(1.84^{+2.86}_{-0.12})^{+0.50}_{-0.04}$ and 
$S_{\rm 1keV}$=$(52.4^{+42.9}_{-42.4})^{+29.0}_{-0}~ \rm nJy$, respectively.
The flux density at 1 keV of the PL from the western lobe indicates a dimmer 
non-thermal component than that of the eastern lobe. Although the western lobe 
is closer to us, the distance between the eastern and western lobes 
($\sim$126 kpc) is negligible in comparison with the distance between Earth 
and Cygnus A ($\sim250$ Mpc). Furthermore, assuming the hotspot speed to be 
the upper limit of the expanding speed of the lobes, the effect of beaming can
be disregarded since the speed of the hotspot is $\sim0.01c $ (Carilli et al. 1991).

Therefore, the differences are possibly attributable to intrinsic differences
 between the individual lobes.

%%%%%%%%%%%%%
% Section 4 %
%%%%%%%%%%%%%
\section{Broadband spectra and physical parameters of the lobes}
In this section,
we estimate the physical parameters of the Cygnus A lobes
by comparing the radio and X-ray data
with those predicted by one-zone emission models.
Radio data analysis for the estimation of 
lobe flux is presented in $\S~4.1$.
The comparisons of the observations and models 
and the obtained physical parameters are presented in $\S~4.2$.

\subsection{Estimation of radio-lobe flux}
We used archived data on Cygnus A at 1.3, 1.7, 4.5, and 5.0 GHz that was 
acquired by the VLA\footnote {The Very Large Array is a facility of the 
National Radio Astronomy Observatory (NRAO),
National Science Foundation.} in various configurations. The A and C VLA configurations were chosen to sample the $u$-$v$ plane adequately on the 
shortest baselines necessary and to provide good resolution when mapping 
the detailed source structure. These data are summarized in Table 
\ref{radiobslog2} and Table \ref{radflux}. The data were calibrated by 
using the Astronomical Image Processing System (AIPS) software package 
developed by the NRAO. 3C~286 was used as a primary flux calibrator. 
Observations in different VLA configurations were imaged separately by 
using the \texttt{CLEAN} algorithm and self-calibrated several times. 
This self-calibration was performed with the Difmap software package 
(Shepherd et al. 1994).  Then, the data sets acquired in different 
configurations for the same frequency were combined to improve the
$u$-$v$ coverage by using the AIPS task \texttt{DBCON}. 
The combined data sets were used to produce the final images after 
a number of iterations with \texttt{CLEAN} and self-calibration with Difmap.

\begin{figure}[h]
\centerline{
\includegraphics[angle=0,width=8cm]{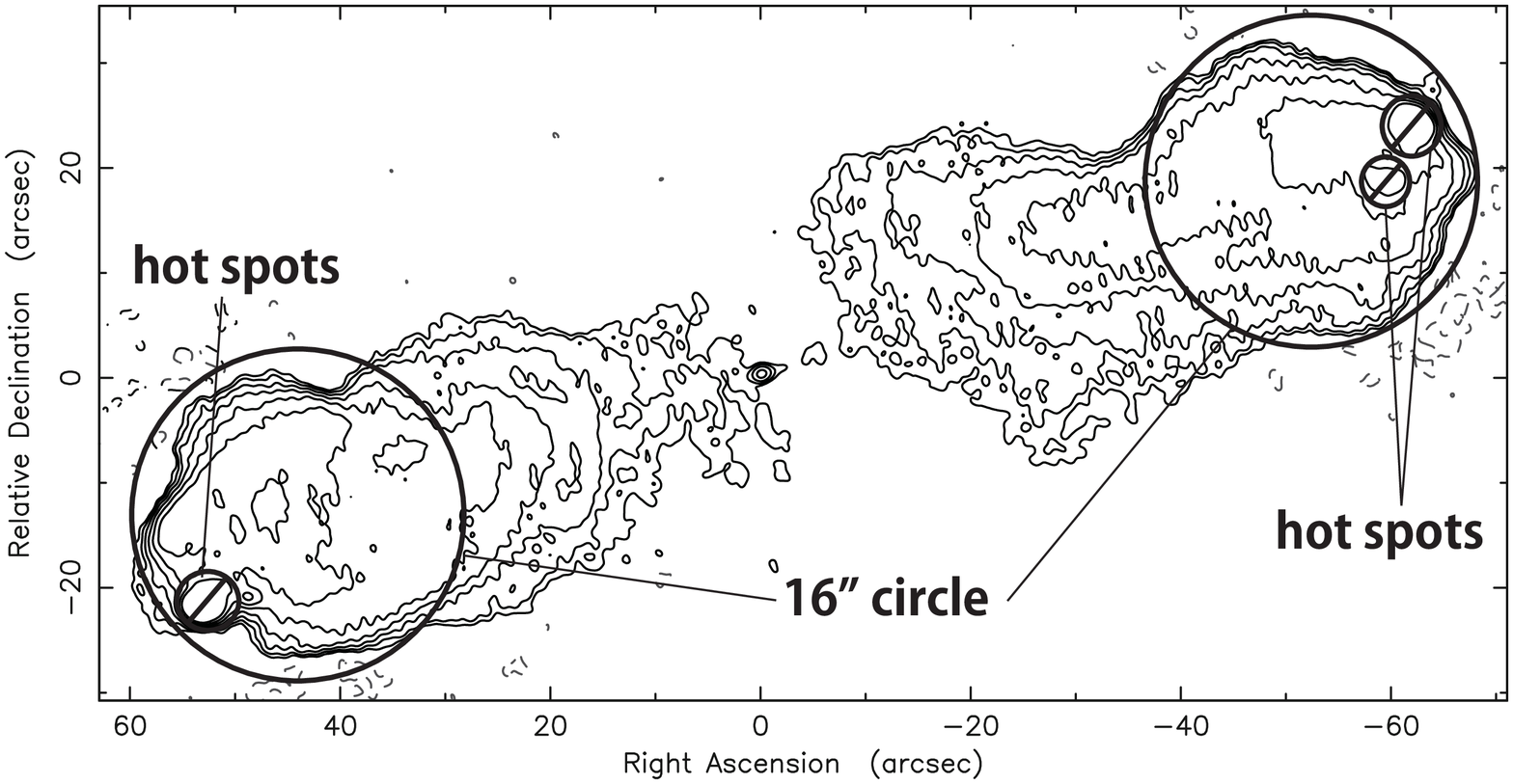}}
\caption{Image of Cygnus A at 1.3 GHz. Contours are plotted at levels of 
3$\sigma$ $\times$ (-1, 1, 2, 4, 8, 16, 32, 64).  
The r.m.s. noise level is $\sigma$=17.53 mJy/beam.
The circle indicates the region where the lobe flux density was estimated.}
\label{radlobfig} 
\end{figure}

Firstly, we convolved the images at all frequencies to obtain the same 
resolution at 1.3~GHz.  Then, we estimated the lobe flux density by 
integrating the flux density within the circle with the radius of 
16$\arcsec$ (hereafter 16$\arcsec$ circle) shown in Figure \ref{radlobfig} 
and subtracting the flux of hotspots from the integrated flux of the
 16$\arcsec$ circle. In the X-ray image, one hotspot and two hotspots 
 can be seen in the eastern and western lobes, respectively (Fig. \ref{X-rayImage_IntgrationRegion}). In the radio images, these hotspots can be 
 clearly identified. We characterized the flux of each hotspot by fitting 
 a model of the emission from a two-dimensional elliptical Gaussian. 
 The estimated hotspot and lobe fluxes are presented in Table \ref{radlobe}. Following the analysis in X-ray, we do not subtract the overlapping lobe 
 fluxes from the Gaussian fit regions\footnote{In the previous works of
 Carilli et al. (1991) and Steenburge et al. (2010), the hotspot fluxes 
 are defined after subtracting the surrounding lobe fluxes contributions. 
 Therefore, our hotspot fluxes in Table 6 are larger than the ones shown 
 in Carilli et al. (1991) and Steenburge et al. (2010).}. We neglected 
 the emission from the jet because the jet emission only accounts for 
 $\sim 5\%$ of the lobe flux (Steenbrugge et al. 2010). Neglecting the 
 emission from the jet does not affect the result of spectral energy 
 distribution (SED) fit in $\S~4.2$ significantly.

%%--------%%
%% Table4 %%
%%--------%%
\begin{deluxetable}{lccccc}
\tablecaption{Properties of VLA data \label{radiobslog2}}
\tablewidth{0pt}
\tablehead{
\colhead{Date }	 &
\colhead{Array}	 &
\colhead{Frequency}	 &
\colhead{Bandwidth}\\	 
 & configuration
 &  (MHz)
 & (MHz)
}
\startdata
1983 Oct 24 & A & 4525, 4995 & 25\\
1984 Apr 15 & C & 4525, 4995 & 25\\
1986 Dec 1 & C & 1345, 1704 & 6.25, 6.25\\
1987 Aug 18 & A & 1345, 1704 & 3.13
\enddata
\end{deluxetable}

%%--------%%
%% Table5 %%
%%--------%%
\begin{deluxetable}{lcccc}
\tablecaption{Properties of VLA data \label{radflux}}
\tablewidth{0pt}
\tablehead{
\colhead{Frequency}	 &
\colhead{Flux}	 &
\colhead{Beam size}	 &
\colhead{P.A.\tablenotemark{a}}	 &
\colhead{$\sigma_{r.m.s.}$\tablenotemark{b}} \\
%\colhead{D.R.\tablenotemark{c}} \\
 (MHz) 
 & (Jy)
 & (arcsec) 
 & (deg)
 & (mJy beam$^{-1}$)
}
\startdata
1345 & 1580.48 & 1.19 $\times$ 1.12 & ~87.6 & 17.53 \\%& 2014\\
1704 & 1290.10 & 0.96 $\times$ 0.91 & ~81.4 & 11.54 \\%& 2062\\
4525 & ~406.58 & 0.39 $\times$ 0.33 & -86.6 &~0.82 \\%&3439\\
4995 & ~366.08 & 0.36 $\times$ 0.30 & -86.6 &~0.83 %& 2717
\enddata
\tablenotetext{a}{Beam position angle.}
\tablenotetext{b}{Rms noise of the image. } 
%\tablenotetext{c}{Dynamic range.}
\end{deluxetable}

%%==========%%
%% Figure 4 %%
%%==========%%

%%--------%%
%% Table6 %%
%%--------%%
\begin{deluxetable}{ccccc}[h]
\tablecaption{Radio lobe / hotspot flux density\label{radlobe}}
\tablewidth{0pt}
\tablehead{
\colhead{Frequency }	 &
\colhead{}	 &
\colhead{$I_{\rm tot} \tablenotemark{a}$}	 &
\colhead{${I_{\rm hot} \tablenotemark{b}}$}	 &
\colhead{${I_{\rm lobe} \tablenotemark{c}}$} \\
(MHz)  &
 & (Jy) 
 & (Jy) 
 & (Jy)
}
\startdata
1345 & Eastern lobe & 718$\pm36$  & 124$\pm6$ & 594$\pm36$\\
	 & Western lobe &  579$\pm29$ &150$\pm5$ & 429$\pm29$ \\
1704 & Eastern lobe & 580$\pm29$ & 117$\pm4$ & 463$\pm29$\\
	 & Western lobe &  477$\pm24$ &120$\pm4$& 357$\pm24$\\
4525 & Eastern Lobe & 192$\pm10$ & 63$\pm2$& 129$\pm10$\\
          & Western lobe &  175$\pm9$  & 53$\pm2$ & 122$\pm9$\\
4995 & Eastern lobe &  174 $\pm9$  & 59$\pm2$  & 115$\pm9$ \\
	 & Western lobe & 159 $\pm8$ & 51$\pm2$ & 108$\pm8$
\enddata
\tablenotetext{a}{$I_{tot}$ corresponds to the total lobe flux within the 16$''$ circle.
The error is the root sum square of flux calibration error (5\%) and thermal noise.}
\tablenotetext{b}{The total flux at hotspots is denoted as $I_{hot}$. 
The error is estimated to be the root sum square of flux calibration error (5\%), gaussian fit error, and thermal noise. }
\tablenotetext{c}{Net lobe flux is evaluated as $I_{lobe}=I_{tot}-I_{hot}$. The error is the root sum square of the error in $I_{tot}$ and the error in $I_{hot}$.}
\end{deluxetable}

%%==========%%
%% Figure 5 %%
%%==========%%
\begin{figure*}[t]
\centerline{
\includegraphics[angle=0,width=8cm]{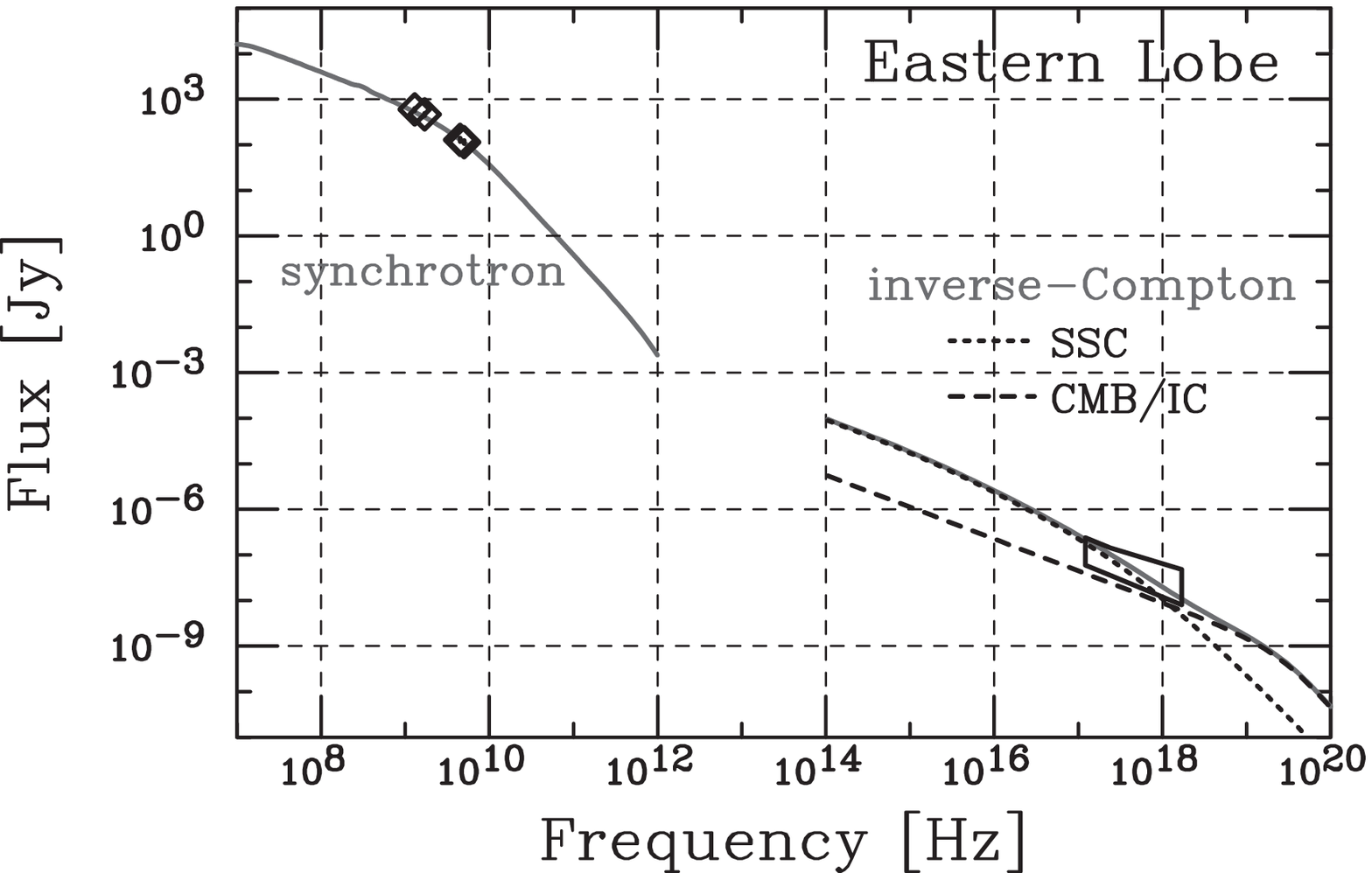}
\includegraphics[angle=0,width=8cm]{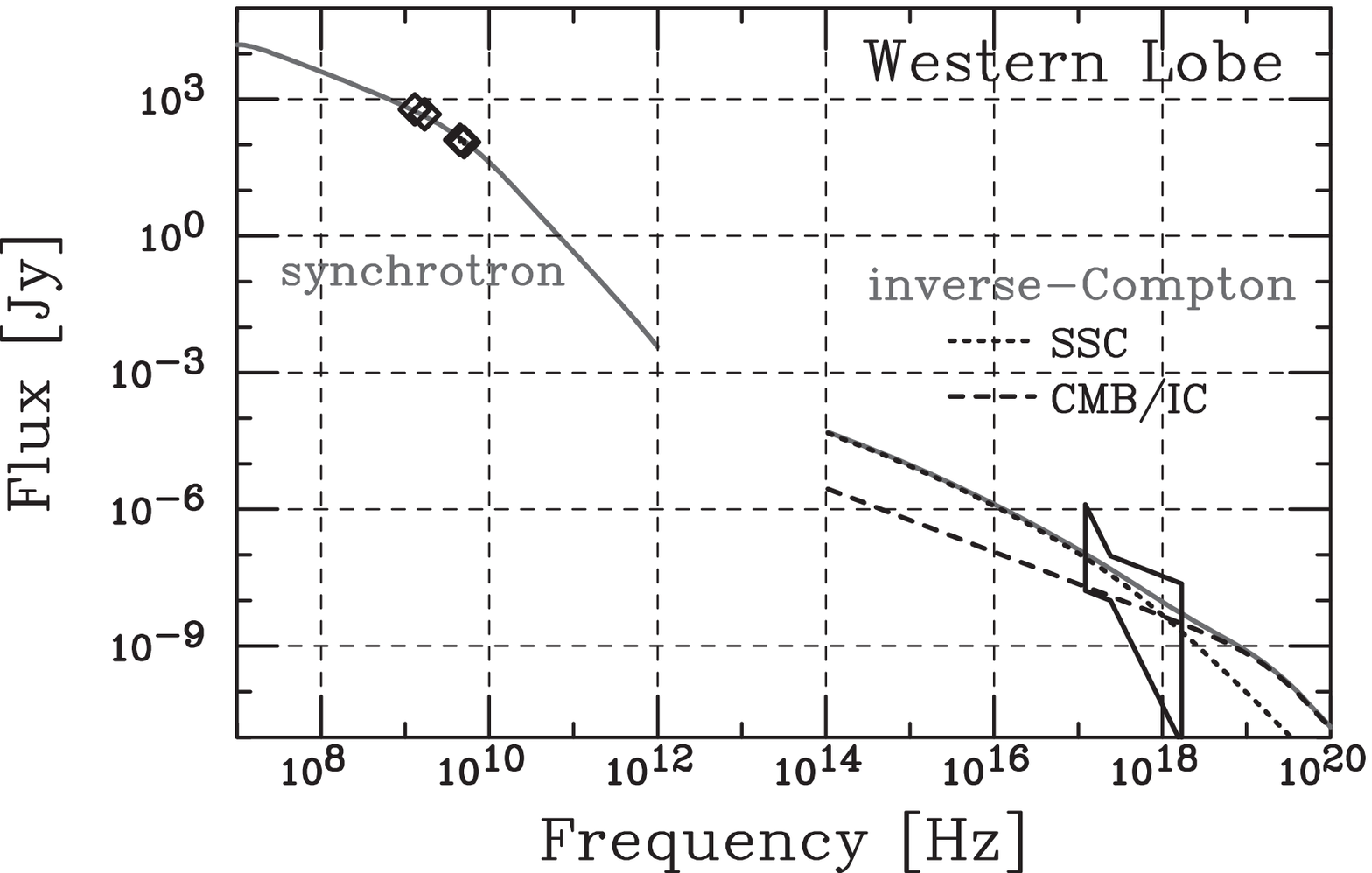}}
\caption{
Spectral energy distribution of the eastern 
(left) and western (right) lobes of Cygnus A.
Areas enclosed by solid lines: X-ray spectrum obtained with $\it Chandra$. 
Diamonds: integration flux obtained at 
1.3 GHz, 1.7 GHz, 4.5 GHz and 5 GHz.
Derived models for synchrotron and IC emissions are shown
by solid lines at lower and higher frequencies.
For the IC emissions, the dashed line represents CMB-boosted IC emission and 
the dotted line represents 
SSC emission. 
}
\label{SED}
\end{figure*}

%%--------%%
%% Table7 %%
%%--------%%
\begin{deluxetable*}{lcccc}
\tablecaption{Physical parameters for lobes of Cygnus A \label{lobPhysical}}
\tablewidth{0pt}
\tablehead{
\colhead{Parameter}	 &
\colhead{~~~~~~~~~~~~~~~~}	 &
\colhead{Unit }	 &
\colhead{Eastern lobe~~~~~~~~}	 &
\colhead{Western lobe}	 
}
\startdata
Volume of lobe (V) &&$\rm cm^{3}$          & \multicolumn{2}{c}{5.8~$\times 10^{68}$}\\
Min electron Lorentz factor ($\gamma_{\rm min}$)  &&--  & \multicolumn{2}{c}{1} \\
Max electron Lorentz factor ($\gamma_{\rm max}$) & &-- & \multicolumn{2}{c}{$10^{5}$}     \\
Normalization of electron energy spectrum ($N_{\rm0}$) && $\rm cm^{-3}$ & 3.1$\times 10^{-3}$ & 1.6$\times 10^{-3}$\\
Electron break Lorentz factor ($\gamma_{\rm b}$) && -- &7$\times 10^{3}$  & 6$\times 10^{3}$ \\
Magnetic field ($B_{\rm IC}$) && $\mu \rm G$        &  15   &  22              \\
Ratio of $B_{\rm IC}$ and $B_{\rm eq}\tablenotemark{a}$ ~($B_{\rm IC}/B_{\rm eq}$)  &&
   --   &  0.30 & 0.44                  \\
Magnetic energy density ($u_{\rm m}$) & & $\rm erg~cm^{-3}$ & $9.0\times 10^{-12}$ & $2.0\times 10^{-11}$      \\
Electron energy density ($u_{\rm e}$) & &$\rm erg~cm^{-3}$  &  $6.0\times 10^{-9}$ & $3.2\times 10^{-9}$             \\
Ratio of $u_{\rm e}$ and $u_{\rm m}$ ($u_{\rm e}$/$u_{\rm m}$) &&
     --      & 666~(35\tablenotemark{b}~)  & 160~(11\tablenotemark{b}~)   
\enddata
\tablenotetext{a}{$B_{\rm eq}=50\mu \rm G$ by Calliri et al. (1991),
calculated under the minimum energy condition.}
\tablenotetext{b}{Values when $\gamma_{\rm min}=1000$.}
\end{deluxetable*}

\subsection{Physical parameters of the lobes}
Figure \ref{SED} shows the X-ray and radio SEDs of  
the eastern and western lobes of Cygnus A. In the X-ray 
analysis, the eastern jet region was excluded from the integrated lobe region 
in a circle with a radius of 16$\arcsec$. In order to match the area fitted 
in the X-ray analysis to that fitted in the radio analysis,
we renormalized the X-ray fluxes by a factor of 1.34 for the eastern lobe.
Consequently, it can be clearly seen that the radio spectrum does not connect 
smoothly to the X-ray spectrum; therefore, diffuse X-rays are produced via 
the IC process caused by SR electrons in the lobe.

\subsection{Physical parameters of the lobes}
In order to determine the origin of seed photons
boosted to the X-ray range, we estimate the energy density
of IR photons ($u_{\rm IR}$), SR photons ($u_{\rm SR}$) and CMB 
photons ($u_{\rm CMB}$) in the lobe. Here, we adopted 
$u_{\rm CMB}=4.1\times 10^{-13}(1+z)^4~\rm erg~cm^{-3}$ 
( Harris \& Grindlay 1979),
$u_{\rm IR}=L_{\rm IR}/4 \pi c d^2~\rm erg~cm^{-3}$,
and assumed IR luminosity of 
$L_{\rm IR} \sim 10L_{\rm X} \sim 1.8 \times 10^{44}~\rm erg~s^{-1}$,
considering the typical spectrum of quasars 
(Sanders et al. 1989 and Young et al. 2002). 
$d$ is the distance from the nucleus (e.g., Brunetti et al. 2001),
and $u_{\rm SR}=3L_{\rm SR}/4 \pi c r^2~\rm erg~cm^{-3}$,
where $L_{\rm SR}$ is the SR luminosity and $r$ is the radius of the lobe.
$L_{\rm SR}$ was calculated by integrating the flux between 10 MHz and 5 GHz,
where we assumed a PL spectrum, $S_{\nu}=S_{\nu_0}(\nu/\nu_{0})^{-\alpha}$ 
with $\alpha$=0.7 (Carilli et al. 1991).
We used the 1.3 GHz flux listed in Table \ref{radlobe} for $S_{\nu_0}$,
and the radius of each lobes was set to 16.8 kpc.
Thus we obtained 
$u_{\rm CMB}\sim 5.6 \times 10^{-13}~\rm erg~cm^{-3}$,
$u_{\rm IR} \sim 2.0 \times 10^{-13}~\rm erg~cm^{-3}$ and
$u_{\rm SR} \sim 7.0 \times 10^{-13}~\rm erg~cm^{-3}$
, respectively. 
Therefore, we consider $u_{\rm CMB}$ and $u_{\rm SR}$ 
to be dominant in the lobe in the following discussion.
In order to estimate the $u_{\rm e}$ and $u_{\rm m}$ 
values that are spatially averaged over the lobes
in the ``E--L'' and ``W--L'' regions, 
the X-ray and radio fluxes were evaluated through modeling.
We used the SR and synchrotron self-Compton (SSC) model 
with software developed by Kataoka (2000) 
and a CMB boosted IC (CMB/IC) component calculated 
in accordance with Harris \& Grindlay (1979).
Here, we assume that SR and IC emissions are produced 
by the same population of relativistic electrons.
According to Carilli et al. (1991), the sharp cutoff in the SR 
radio spectrum indicates a broken power law with an energy index 
before and after the breaks at 0.7 and 2. We assume the electron
distribution is $N_{\rm e}(\gamma) = N_{0} \gamma^{-\rm s_1}$ 
for $ \gamma<\gamma_{\rm b}$ and 
$N_{\rm e}(\gamma) = N_{0}\gamma_{\rm b}^{s_2-s_1} \gamma^{-\rm s_2}$
for $\gamma> \gamma_{\rm b}$,
where $N_0$ is the normalization of the electron energy spectrum,
$\gamma$ is the electron Lorentz factor and 
$s$ is the electron energy spectrum index.
Moreover, $\gamma_{\rm b}$ is the break Lorentz factor.  
The index of $s_1$ and $s_2$ is 2.4 and 5 ($s=2\alpha+1$), respectively, 
and the assumed range of $\gamma_{\rm min}-\gamma_{\rm max}$ is $1-10^{5}$. 
We fixed the volume parameter to 
$V$=$4\pi r^3 /3$=$5.8 \times 10^{68}~\rm cm^{3}$ of the lobe
and assumed a sphere with a radius of $16.8~\rm kpc$,
which was the same as the radius 
of the integration circle in the X-ray and radio analyses (\S~3). 
We tuned the magnetic field $B$, $\gamma_{\rm b}$ and $N_{0}$.
From the eastern and western radio spectra, the SR turnover 
frequency is assumed to be around 1 GHz.
Under the above conditions, we evaluated the appropriate SR, 
SSC and CMB/IC components to reproduce the X-ray and radio fluxes, 
which are plotted in Fig. \ref{SED}.
The X-ray data is reproduced well by the CMB/IC and SSC emissions. 
Unlike the emissions from other radio lobes that have been reported 
to date (\S~1), the SSC component in the emissions from the lobes of
Cygnus A contributed significantly to the X-ray band.
The SSC component accounts for 80\% and 40\% of the total IC component 
at $\sim$0.6 keV and $\sim$ 7 keV, respectively.
The SSC X-ray component is produced by electrons with a Lorentz factor of 
$\sim$$10^{4}$--$10^{5}$, 
and  SR emission in the $\sim$1--$10^2$ GHz band is produced.
Therefore, it is natural to assume that the same distribution of 
electrons produces the radio and X-ray emissions.
The derived parameters are summarized in Table \ref{lobPhysical}.
The ratio $B_{\rm IC}/B_{\rm eq}=0.30$ and $0.44$ for the eastern and 
western lobes, respectively, are in the range previously reported for 
other objects, $B_{\rm IC}/B_{\rm eq}=(0.1-1)$ (e.q., Croston et al. 2005).
The ratio $u_{\rm e}/u_{\rm m}$ appears to show significant electron 
dominance in the lobes of Cygnus A.
Taking $\gamma_{\rm min}=1000$ in order to make a comparison with other 
radio galaxies, we obtain $u_{\rm e}/u_{\rm m}=$ 35 and 11 for the eastern 
and western lobes, respectively.
Thus, we found that $u_{\rm e}/u_{\rm m}$ for the lobes of Cygnus A 
at the center of the cluster is similar to that for other field radio 
galaxies (\S~1).

%%%%%%%%%%%%%
% Section 5 %
%%%%%%%%%%%%%
\section{Summary}
Using $\it Chandra$ deep observation data (230 ks) for Cygnus A,
we carefully analyzed the X-ray spectra of the lobes and the 
regions surrounding the lobes. Our findings are as follows. 

\begin{itemize}

\item
In $\it Chandra$ X-ray images, among emissions originating from ICM,
we confirmed extended X-ray emission regions corresponding to the 
eastern and western lobes.

\item
The X-ray spectra of the lobe regions could not be reproduced 
by a single Mekal model, and we found that the addition of a PL 
component was more appropriate than the addition of an additional 
Mekal component in the statistical analysis. The best-fit photon 
indices of the eastern and western lobe regions were $1.69^{+0.07}_{-0.13}$
and $1.84^{+2.90}_{-0.12}$, and the flux densities at 1 keV were 
$77.7^{+28.9}_{-31.9}~\rm nJy$ and $52.4^{+42.9}_{-42.4}~\rm nJy$, 
respectively.

\item
The obtained X-ray and radio SED of the lobes supported 
the IC mechanism for X-ray emission. Furthermore, the 
X-rays are likely produced via both SSC processes below $\sim 10^{18}$ Hz 
(4 keV) and CMB/IC processes above $\sim10^{18}$ Hz (4 keV). This is the 
first case of a lobe where SSC emission has been found to affect IC emission.

\item
The derived physical parameters under the SSC model indicate that
the energy density of electrons dominates that of magnetic fields
both in the eastern and western lobes, 
as often reported from other radio lobe objects.

\end{itemize}

\acknowledgments
We would like to appreciate the referee for his/her careful reading
and constructive comments to improve this paper. We have made use of the VLA
archive data.The VLA is operated by the National Radio Astronomy Observatory, 
a facility of the National Science Foundation (NSF) operated under
cooperative agreement by Associated Universities, Inc.

\end{document}